# Ultrafast response of plasma-like reflectivity edge in (TMTTF)$_2$AsF$_6$ driven by 7-fs, 1.5-cycle strong-light field

Y. Naitoh[1], Y. Kawakami[1], T. Ishikawa[1], Y. Sagae[1], H. Itoh[1], K. Yamamoto[2],

T. Sasaki[3], M. Dressel[4], S. Ishihara[1], Y. Tanaka[5], K. Yonemitsu[5], and S. Iwai[1*]

[1]Department of Physics, Tohoku University, Sendai 980-8578, Japan

[2]Department of Applied Physics, Okayama Science University, Okayama, 700-0005

Japan

[3]Institute for Materials Research, Tohoku University, Sendai 980-8577, Japan

[4] 1. Physikalisches Institut, Universität Stuttgart, 70569, Stuttgart, Germany

[5]Department of Physics, Chuo University, Tokyo 112-8551, Japan

71.27.+a, 75.30.Wx, 78.47.jg

*s-iwai@m.tohoku.ac.jp

Abstract

The strong light-field effect of (TMTTF)$_2$AsF$_6$ was investigated utilizing 1.5-cycle, 7-fs infrared pulses. The ultarfast (~20 fs) and large (~40%) response of the plasma-like reflectivity edge (~0.7 eV) was analyzed by the changes in $\omega_p = \sqrt{ne^2/(\varepsilon_\infty \varepsilon_0 m)}$ ($n$: number of charges in the 1/4 filled-band, $m$: mass of charge, $\varepsilon_\infty, \varepsilon_0$: dielectric constants for high-frequency and vacuum, $e$: elementary charge). The 3% reduction in $\omega_p$ is attributed to the 6% increase in $m$. Furthermore, 20 fs oscillation of $\omega_p$ in the time domain indicates that the plasma-like edge is affected by the charge gap (~0.2 eV) nature. Theoretical calculations suggest that the Coulomb repulsion plays an important role in the increase in $m$.

I. Introduction

The stimulation of strongly correlated systems by light pulses enables dramatic and ultrafast changes in the electronic phase[1-3]. Furthermore, in the framework of the dynamical mean-field theory (DMFT)[4] and others [5-9], strong field (> 1 MV/cm) effects of correlated charges have been predicted, such as dynamical localization (reduction of the transfer integral $t$ or increase in mass of charge $m$), negative temperature, and repulsive-attractive conversion in the Coulomb interaction. Recent ultrafast measurements utilizing <10-fs pulses can access such nonequilibrium electronic states with the lifetime of several tens of femtoseconds as an "optical freezing" of the charge motion in an organic conductor [10]. The mechanism of such a light-field-induced metal-to-insulator transition has been discussed in terms of the increase in $m$ (or decrease in $t$), although the direct detection of $m$ in a short time scale has not yet been performed.

The tetramethyltetrathiafulvalene (TMTTF) compound $(TMTTF)_2AsF_6$ is a typical quarter-filled organic conductor [11-16], exhibiting charge ordering (CO) below $T_{CO}$=102 K, although the correlation effect in this compound is rather weak. Such a small-gap (0.1–0.2 eV) insulator has a near-infrared (~0.7 eV) reflectivity edge that is analogous to the plasma edge of metals [12, 15, 17]. This plasma-like reflectivity edge can be characterized by $\omega_p = \sqrt{ne^2/(\varepsilon_\infty \varepsilon_0 m)}$ in the Lorentz model, if $\omega_p \gg \omega_0$ (the number of charges $n$：~2 x $10^{21}$cm$^{-3}$ in the 1/4 filled-band and their mass $m$：3–4$m_0$, dielectric constants for high-frequency and vacuum $\varepsilon_\infty, \varepsilon_0$, charge gap $\hbar\omega_0$：0.1–0.2 eV).

In this sense, we refer to this reflectivity edge as a "plasma-like edge", because it reflects the collective charge excitation much above the gap. Therefore, in this compound, we can expect to observe the field-induced change in $m$ or $t$ by measuring the spectral change in the plasma-like edge around $\hbar\omega_p$ ~0.7 eV.

In this study, we demonstrate ultrafast (~20 fs) increase in $m$ of ~6% in (TMTTF)$_2$AsF$_6$ under application of 1.5 cycle and 9.8 MV/cm electric field, by detecting a ~3% decrease in $\omega_p$. The coherent oscillation of $\omega_p$ in the time domain with frequency $\omega_0$ indicates that the plasma-like edge is attributed to the collective charge excitation much above the gap. Furthermore, theoretical considerations indicate that the Coulomb repulsion contribution to the reduction of $\omega_p$ is very important.

## II. Experimental

The steady state reflectivity spectrum was recorded by a Fourier-transform IR spectrometer, Nicolet Nexus 870, equipped with an IR microscope, Spectra-Tech IR-Plan. Transient reflectivity measurements were performed by using both 100-fs and 7-fs pulses. The central photon energy of the pump light for the 7-fs measurement (~0.8 eV) is almost equal to that of the 100-fs measurement (0.89 eV). The probe ranges for the two measurements are 0.1–1 eV (100 fs) and 0.58–0.95 eV(7 fs), respectively. The broadband spectrum for the 7 fs pulse covering 1.2–2.3 μm is shown by the orange shaded area in Fig. 1(a). This spectrum was obtained by focusing a carrier-envelope phase (CEP) stabilized idler pulse (1.7 μm) from an optical parametric amplifier (Quantronix HE-TOPAS pumped by Spectra-Physics Spitfire-Ace) onto a hollow fiber set within a Kr-filled chamber (Femtolasers). Pulse compression was performed using both active mirror and chirped mirror techniques. The pulse width derived from the autocorrelation was 7 fs, which corresponds to 1.5 optical cycles. In the transient reflectivity measurements using the 7-fs pulse, the instantaneous electric field on the sample surface (excitation diameter of 200 μm) can be evaluated as $E_{peak} = 9.8 \times 10^6 (\text{V/cm})$ for an excitation intensity $I_{ex}$ of 0.9 mJ/cm$^2$. Polarized optical reflection experiments were conducted on a single crystal of (TMTTF)$_2$AsF$_6$ (of size 1.5 × 0.7 × 0.2 mm) grown by electro-crystallization [14].

### III. Steady state reflectivity

Figures 1(a) and 1(b) show the optical conductivity and reflectivity (R) spectra of (TMTTF)$_2$AsF$_6$ at 25 and 150 K for the polarization $E//a$, where $a$ is the stacking axis of the planar molecules. The reflectivity spectra [Fig. 1(b)] have a plasma-like edge even below $T_{CO}$ because $\omega_p >> \omega_0$, and the spectra at > 0.2 eV can be well reproduced using the Lorentz model (solid lines) by considering vibrational coupling [12, 18]. The disagreement for vibrational peaks (0.1–0.2 eV) at 25 K may be attributable to the dielectric screening of the vibrational response by the charge [18]. The fitting parameters are $\hbar\omega_p = 0.703$ eV with the scattering rate $\gamma = 0.125$ eV(25 K) and 0.154 eV(150

K) and the charge gap $\hbar\omega_0$ = 0.180(25 K) and 0.193 eV(150 K). All these parameters [18] are roughly equal to those of (TMTTF)$_2$PF$_6$ at 300 K in the previous study [12]. The spectral shape around the reflectivity edge (~0.7 eV) is governed by $\omega_p$ and $\gamma$, i.e., the other parameters describing the charge gap nature and the vibrational responses [12, 18] in the low energy (< 0.2 eV) spectrum do not considerably affect the spectrum around the plasma-like edge.

The inset in Fig. 1(b) shows the temperature dependence of $\gamma$ as $\Delta\gamma/\gamma_{25K}=(\gamma-\gamma_{25K})/\gamma_{25K}$. It is noteworthy that $\gamma$ is almost independent of the temperature below $T_{CO}$, but it abruptly starts to increase at $T_{CO}$ with increasing temperature. The parameter $\gamma$ in the Lorentz model reflects the scattering rate as well as the density of states, especially for one-dimensional systems where the van-Hove singularity appears. However, in this case, the anomalous temperature dependence of $\gamma$ can be reasonably explained by the temperature dependence of the electron–electron scattering, i.e., the frozen charges at $T < T_{CO}$ cannot increase the scattering rate with increasing temperature, although the scattering of the mobile charge should be increased at $T > T_{CO}$.

## IV. Transient reflectivity measured by 100-fs pulse

Figure 2(a) shows the transient reflectivity ($\Delta R/R$) spectra of (TMTTF)$_2$AsF$_6$ (15 K) for time delay, $t_d$ = 0–4 ps, measured using a 100-fs pulse (excitation intensity $I_{ex}$ = 0.5 mJ/cm$^2$, 2 MV/cm). The polarizations of the pump and the probe pulses ($E_{pu}$ and $E_{pr}$, respectively) are parallel to the *a*-axis ($E_{pu}$ //*a*, $E_{pr}$//*a*). The excitation energy 0.89 eV, where the absorption coefficient is very low, was set for non-resonant strong field application.

The $\Delta R/R$ spectrum at $t_d$ = 0.1 ps shown in Fig. 2(b) is well reproduced by a 1.8 % decrease in $\omega_p$, a 12 % increase in $\gamma$ and a 11 % increase in the width of the vibrational peak at 0.165 eV [18] in the Lorentz model. The spectral change calculated with only decreasing $\omega_p$ (blue line) and that with only increasing $\gamma$ (red line) are shown in Fig. 2(c). We can easily distinguish the contributions from $\Delta\omega_p/\omega_p$ (blue curve) and $\Delta\gamma/\gamma$ (red curve). It is noteworthy that a ~2 % reduction in $\omega_p$ can be detected as a ~30 % change of $\Delta R/R$ in the spectral range of $\omega_p$. On the other hand, the 12 % increase in $\gamma$ indicates that temperature increases up to ~120 K across $T_{CO}$. We can notice a small structure at ~0.12 eV around the vibrational peaks in Fig. 2(b). The corresponding structure is also seen in the calculation (orange curve).

## V. Transient reflectivity measured by 7-fs pulse

Figure 3(a) shows the $\Delta R/R$ spectra, for $t_d$ = 0–150 fs, measured using the 7-fs pulse ($I_{ex}$ = 0.8 mJ/cm$^2$, 9.8 MV/cm$^{-1}$). The spectral window of the probe light 0.58–0.95 eV is shown by the white arrow in Fig. 2(a). The excitation pulse energy and the field amplitude of the 7-fs pulse are higher than those of the 100-fs pulse. Therefore, a quantitative comparison between both results is difficult. However, the spectral shapes of $\Delta R/R$ are qualitatively consistent at $t_d$ = 0.1 ps, although the values of $\Delta\omega_p$ and $\gamma$ are different, as described below. Figure 3(b) shows the time profiles of $\Delta R/R$ measured at i) 0.85, ii) 0.73 and iii) 0.62 eV. As shown in Fig. 3(b) (i), the decrease in $R$ appears in the time scale of ~20 fs, and has an oscillating structure with the period of 20 fs at 0.85 eV, reflecting the ultrafast reduction and the coherent oscillation of $\omega_p$ [19]. On the other hand, the slower rise (~80 fs) of $R$ is observed at 0.62 eV [Fig. 3(b) –(iii)], where $\Delta R/R$ shows an increase in $\gamma$ [Fig. 2(c)].

The $\Delta R/R$ spectra at various time delays $t_d$ = 0–80 fs are shown in Figs. 4(a)–4(f). At $t_d$= 18 fs [Fig. 4(b)], $\Delta R/R$ of about -0.4 at around 0.7 eV can be reproduced by the 2.8% decrease in $\omega_p$ [blue-dashed curve in Fig. 4(b)]. On the other hand, at $t_d$ =80 fs, a 30% increase in $\gamma$ [red-dashed curve in Fig. 4(f)] with a 1.7% decrease in $\omega_p$ [blue-dashed curve in Fig. 4(f)] is required in order to reproduce the spectrum. These spectral changes indicate that the ultrafast (~20 fs) reduction of $\omega_p$ and the slower (<80 fs) increase in $\gamma$ reflecting an electron temperature increase.

In general, the decrease in $\omega_p$ indicates an increase in $m$ or a decrease in $n$. Here, $n$ is the number of the charges in the 1/4 filled-band system with charge 0.5 e per TMTTF molecule on average. Therefore, the number of these charges does not decrease upon excitation by the photon energy well below any interband transition from the present 1/4 filled-band. Since a strong light field of 9.8 MV/cm can reduce $t$ in α-(ET)$_2$I$_3$, the origin of the decrease in $\omega_p$ is also attributable to the increase in *m*. In fact, the 2.8 % decrease in $\omega_p$ (5.8% increase in *m*) is consistent with a ~10% decrease in *t* estimated in α-(ET)$_2$I$_3$ [10].

The Δ*R*/*R* spectra at $t_d$ = 35 and 50 fs [Figs. 4(d) and 4(e)] exhibit a spectral dip at ~0.7 eV, as shown by the green shade in the figure. They cannot be reproduced in the framework of the Lorentz model. We cannot discuss the electron–electron or electron–phonon scatterings in such an early time region by using $\gamma$, i.e., a description using a stochastic process is not valid, because i) scattering occurs only once or twice for $\hbar/\gamma$ of about 40 fs, and ii) the coherence of charges can survive in such early stage. Therefore, instead of a stochastic description, we should employ a deterministic description using an interaction between the oscillating charges with frequency $\omega_p$ and other charges and/or lattice. To discuss the origin of this spectral dip, we should make further investigations from both experimental and theoretical viewpoints. However, they are beyond the scope of the present study.

Figure 5(a) shows the time profiles of $-\Delta\omega_p/\omega_p$ (blue dots) and $\Delta\gamma/\gamma$ (red dots), which were obtained using the Lorentz analysis (shown by the blue and red curves in Fig. 4). The time profile of the spectral area of the dip [green shades in Figs. 4(d) and 4(e)] is shown in Fig. 5(b). The dip appears at $t_d$ = ~40 fs, which is approximately equal to the averaged scattering time $\hbar/\gamma$, although the physical meaning of this agreement remains unclear. The decay time of the dip agrees well with the growth time of $\gamma$, reflecting the crossover from coherent scattering to a stochastic process.

In the time profile of $-\Delta\omega_p/\omega_p$ in Fig. 5(a), an oscillating structure with a period of 20 fs can be observed. This oscillation is also seen in the raw data Figs. 3(a) and 3(b)–i). The oscillation period corresponds to the conductivity peak in Fig. 1(a), reflecting the dimerization gap (~0.2 eV) in the weakly dimerized lattice with $U$ [14, 20]. The different energy scales between $\omega_p$ (0.7 eV) and the charge gap (~0.2 eV) allow us to detect the time-domain oscillation of $\omega_p$ with frequency corresponding to the charge gap. In other words, on the energy scale at 0.7 eV, the behavior of charges is quite similar to those in ordinary metals. Although the near-infrared 7-fs pulse cannot directly access the charge gap of ~0.2 eV, the charge gap affects the plasma-like reflectivity edge at 0.7 eV through the oscillation structure in the time profile.

Here, we discuss the difference between the results of (TMTTF)$_2$AsF$_6$ and those of α-(ET)$_2$I$_3$ [10]. In quasi two-dimensional conductor α-(ET)$_2$I$_3$, the reduction of $t$ just above $T_{CO}$ can induce a metal-to-insulator transition accompanied by the drastic change in the optical conductivity, i.e., spectral weight transfers from lower to higher energy regions. Thus, we easily detect the light-field-induced metal-to-insulator transition. On the other hand, in quasi one-dimensional conductor (TMTTF)$_2$AsF$_6$, quantum fluctuations are more effective to reduce the long-range order. In fact, the charge disproportionation in (TMTTF)$_2$AsF$_6$ [0.585(rich)-0.415(poor)] is much smaller than that in α-(ET)$_2$I$_3$ [0.8(rich)-0.2(poor)]. Thus, excitations in the near-infrared region are little influenced by the charge disproportionation in this compound. Indeed, near-infrared reflectivity at low temperature are well reproduced by the Lorentz model which is characterized by $\omega_p$ and $\gamma$, while the weak bond alternation is responsible for the dimerization gap [20] at ~0.2 eV. Therefore, the near-infrared spectrum of (TMTTF)$_2$AsF$_6$ is suitable for detecting the field-induced increase in $m$ (or decrease in $t$), although this compound is not suitable for seeking any field-induced transition.

It is finally noted that, although both $\omega_p$ and $\omega_0$ (dimerization gap) can be affected by the modulation of $t$, the relation between $t$ and $\omega_0$ (dimerization gap) is quite different from that between $t$ and $\omega_p$, i.e., $\omega_p$ is proportional to $\sqrt{t}$, while $\omega_0$ is proportional to $(t_2 - t_1)$ [20]. The reduction of $\omega_p$ is much more essential because $t_2$ and $t_1$ in the equation for $\omega_0$ are equally affected by the electric field.

## VI. Theoretical consideration using the time-dependent Schrödinger equation

The field-induced increase in $m$ should be theoretically described in a correlated electron system after strong photoexcitation. Thus, we employ the one-dimensional, quarter-filled, weakly dimerized, extended Hubbard model,

$$H = \sum_{i,\sigma} t_{i,i+1}\left(c_{i\sigma}^+ c_{i+1\sigma} + c_{i+1\sigma}^+ c_{i\sigma}\right) + U\sum_i n_{i\uparrow}n_{i\downarrow} + V\sum_i n_i n_{i+1} \quad ,$$

where $c_{i\sigma}$ is the annihilation operator of a hole on the $i$ th site with spin $\sigma$, $n_{i,\sigma} = c_{i\sigma}^+ c_{i\sigma}$, and $n_i = \sum_\sigma n_{i,\sigma}$. This model has on-site ($U$) and nearest-neighbor ($V$) repulsive interactions and alternating transfer integrals ($t_{i,i+1} = t_1, t_2, t_1, t_2, \cdots$). We use an exact diagonalization method for the 16-site chain with the anti-periodic boundary condition, and set $t_1 = 0.16$ eV and $t_2 = 0.2$ eV [21], which were evaluated using first-principles band calculations. By numerically solving the time-dependent Schrödinger equation after the photoirradiation of a monocycle pulse with central frequency $\hbar\omega = 0.7$ eV and $eaF/\hbar\omega = 1$ ($a$: lattice spacing, $F$: field amplitude), we calculate the change in the electronic structures.

We actually calculated six-lattice-spacing-distant off-diagonal density, $-\sum_\sigma \langle c_{0\sigma}^+ c_{6\sigma} + c_{6\sigma}^+ c_{0\sigma} \rangle$, as an index of the delocalized nature of the charges (or equivalently the increase in $m$), by assuming that the molecules are equidistant. We plot the time evolution of this density in Fig. 6. In the noninteracting case of $U=V=0$ (black line), the density simply oscillates and is almost undamped. The oscillation period corresponds to the charge excitation energy $\sim 2\sqrt{t_1^2 + t_2^2}$ for small $U$. For large $U$, a dimerization gap of $\sim 2(t_2 - t_1)$ appears and is dominant [20]. With increasing $U$ and $V$ (red and blue lines), the density is damped or more strongly suppressed. The time average is reduced with increasing field amplitude $F$ and interaction strengths $U$ and $V$ (not shown). In other words, the field-induced suppression of the electronic motion is enhanced by the interactions[22]. The damped oscillation for intermediate ($U/t_2$, $V/t_2$) = (1.0, 0.55) (red line) [23] is similar to the experimentally observed behavior of the plasma frequency. Moreover, the oscillation period shown in Fig. 6 is ~2.5 times shorter than that for the time profile of $\omega_p$ in Fig. 3(b). This comes from the fact that the calculated excitation energy $\sim 2\sqrt{t_1^2 + t_2^2}$ (using $t_1$ and $t_2$ from first-principles band calculations for small $U$) is 2.5 times larger than the observed gap, which is attributed to the dimerization gap $\sim 2(t_2 - t_1)$ for more realistic $U$.

## VII. Summary

A ~3% reduction in $\omega_p$, reflecting a ~6% increase in $m$ is induced by a 9.8 MV/cm instantaneous field in the organic conductor (TMTTF)$_2$AsF$_6$. The coherent modulation of $\omega_p$ with a period of 20 fs indicates that $\omega_p$ is affected by the lower energy gap nature. According to theoretical calculations, the contribution from the Coulomb repulsion plays an important role in the increase in $m$.

[23] The nearest-neighbor repulsive interaction $V$ is regarded as an effective one, to which intrachain and interchain intersite repulsive interactions contribute.

Figure captions

Fig.1 (a) Optical conductivity ($\sigma$) spectra of (TMTTF)$_2$AsF$_6$ at 25 K and 150 K. The spectrum of the 7-fs pulse (orange shaded area) is also shown. (b) Reflectivity ($R$) spectra at 25 K and 150 K with the Lorentz analysis (solid lines). Inset shows the temperature dependence of $\Delta\gamma/\gamma$.

Fig. 2 (a) Time evolution of the transient reflectivity ($\Delta R/R$) at $t_d$ < 4 ps measured by a 100-fs pulse. (b) $\Delta R/R$ spectrum at $t_d$ = 0.1 ps with the Lorentz analysis (solid line). (c) Spectral change calculated with only decreasing $\omega_p$ (1.8%) (blue) and that with only increasing $\gamma$ (12%)(red). A 11% increase in the width of the vibrational peak at 0.165 eV [18] is taken into account.

Fig. 3 (a) Time evolution of $\Delta R/R$ at $t_d$ < 150 fs measured by a 7-fs pulse. (b) Time evolutions of $\Delta R/R$ measured at i) 0.85, ii) 0.73, and iii) 0.62 eV.

Fig. 4 $\Delta R/R$ spectra at various time delays $t_d$ = 0–80 fs are shown as the circles. The blue-dashed, red-dashed, and orange curves indicate the calculated spectral change using the Lorentz model (orange), the spectral change calculated with only $-\Delta\omega_p/\omega_p$ (blue-dashed) and that with only $\Delta\gamma/\gamma$ (red-dashed). The arrows indicate $\hbar\omega_p$.

Fig. 5 (a) Time profiles of $-\Delta\omega_p/\omega_p$ (blue dots) and $\Delta\gamma/\gamma$ (red dots) obtained using Lorentz analysis (shown by the blue-dashed and red-dashed curves in Fig. 4). (b) Time profile of the spectral area of the dip (green shade in Fig. 4).

Fig. 6 Calculated time evolutions of six-lattice-spacing-distant off-diagonal density $-\sum_{\sigma}\langle c^+_{0\sigma}c_{6\sigma} + c^+_{6\sigma}c_{0\sigma}\rangle$ as an index of the delocalized nature of charges for ($U/t_2$, $V/t_2$) = (0, 0) (black), (1.0, 0.55) (red line), and (2.0, 1.1) (blue line). Schematic illustration of the CO molecular stack along the *a*-axis is also shown.

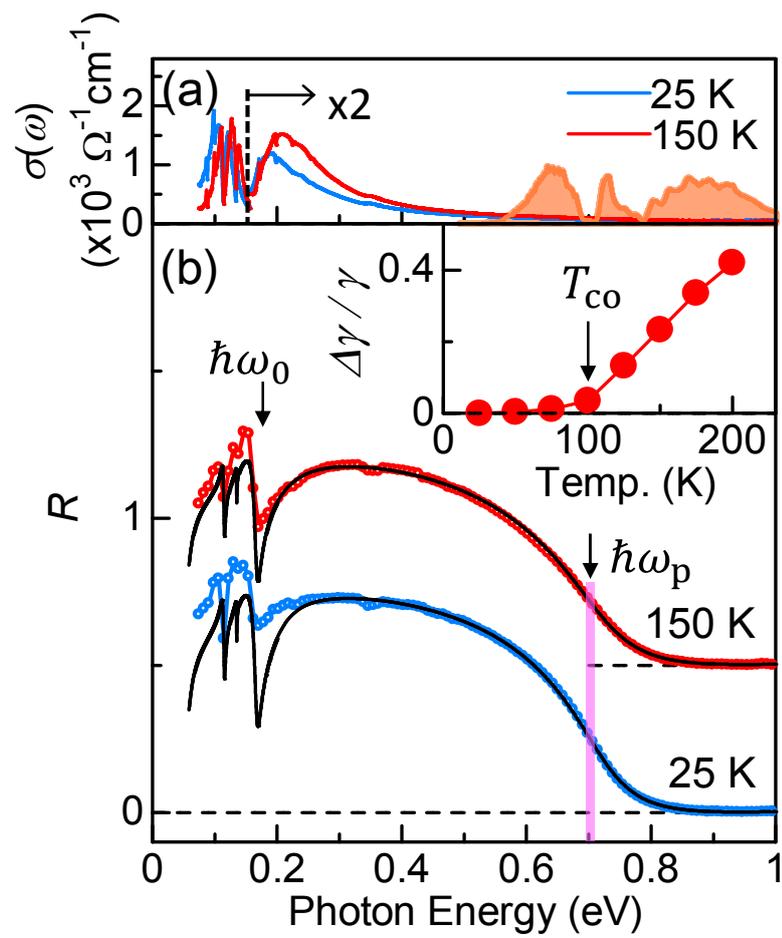

Fig.1

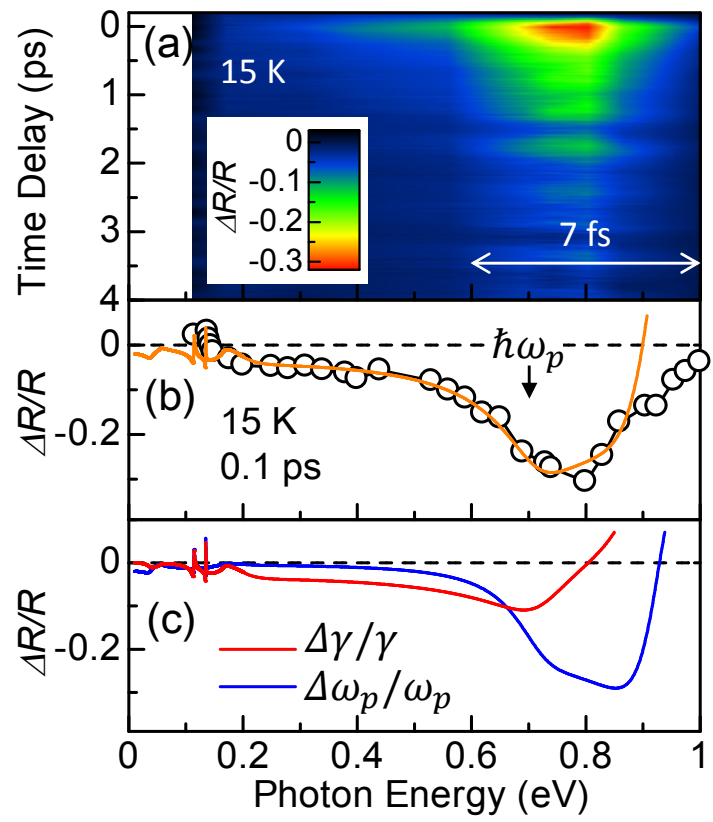

Fig.2

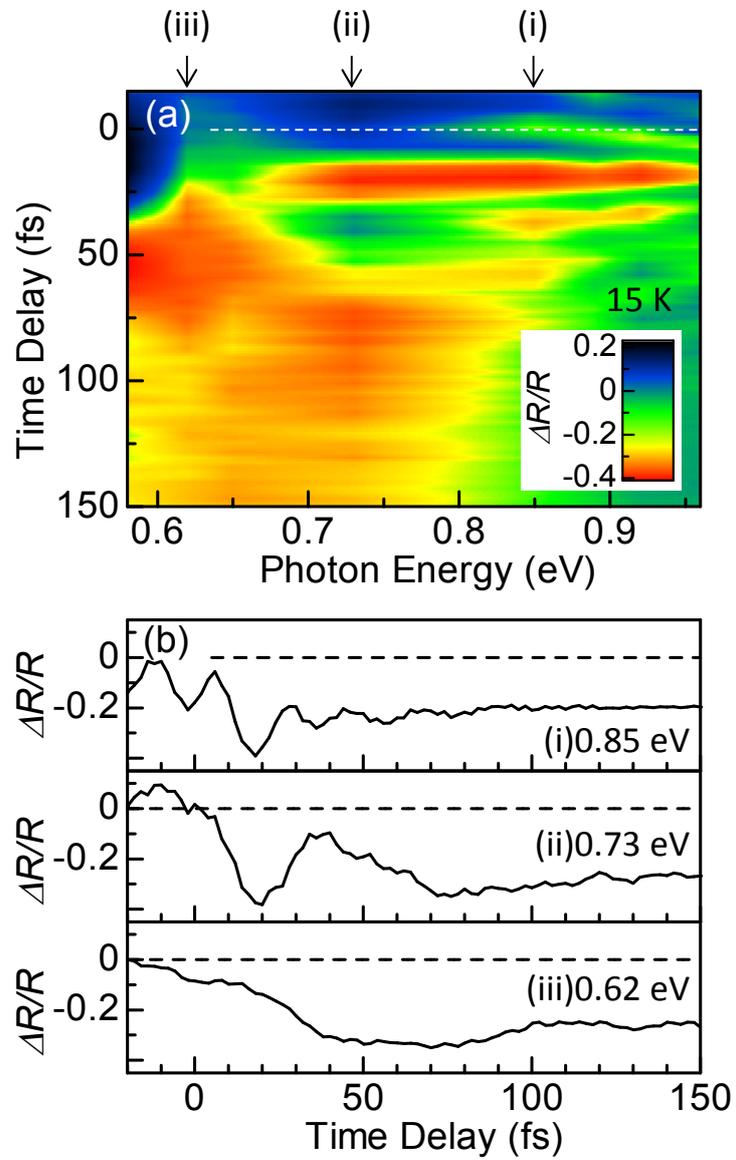

Fig.3

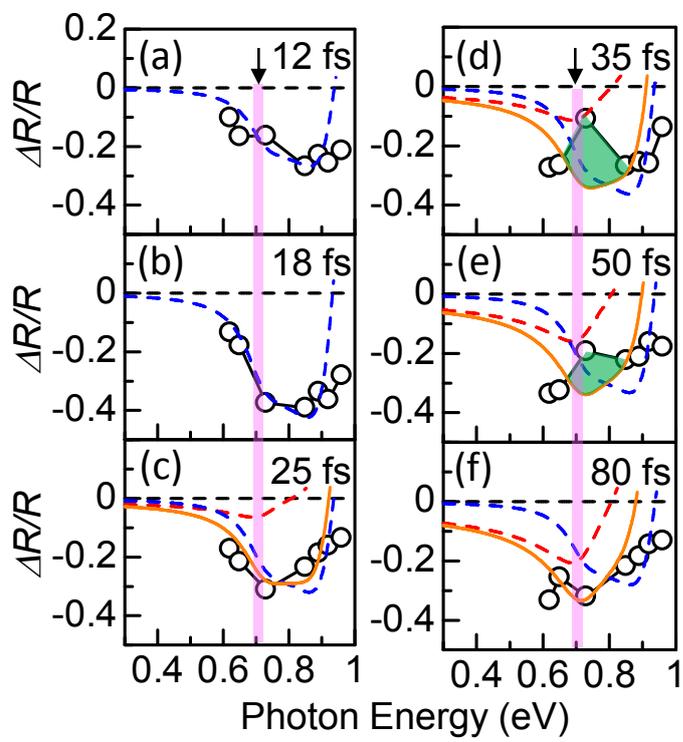

Fig.4

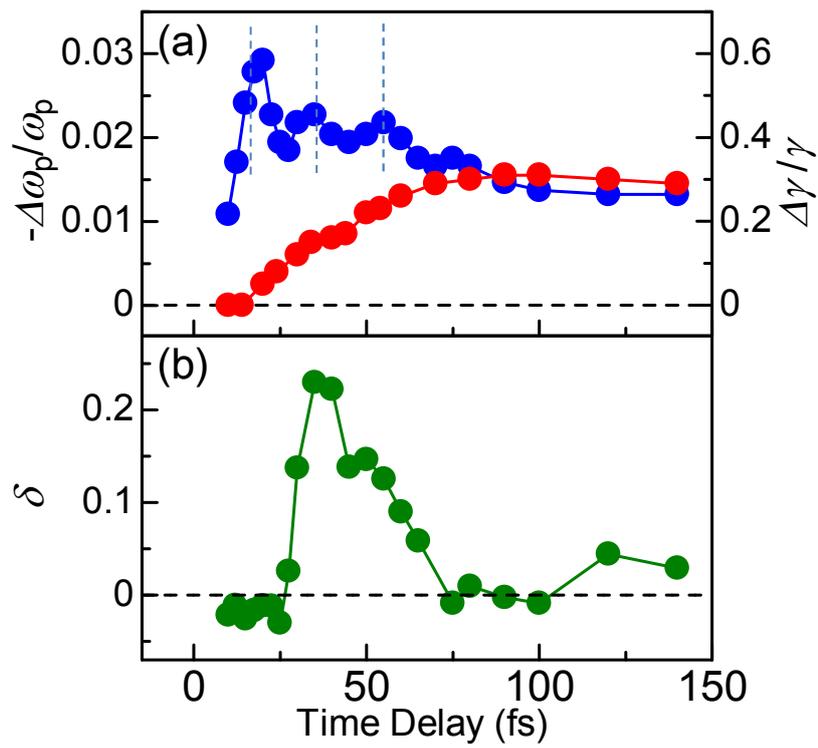

Fig.5

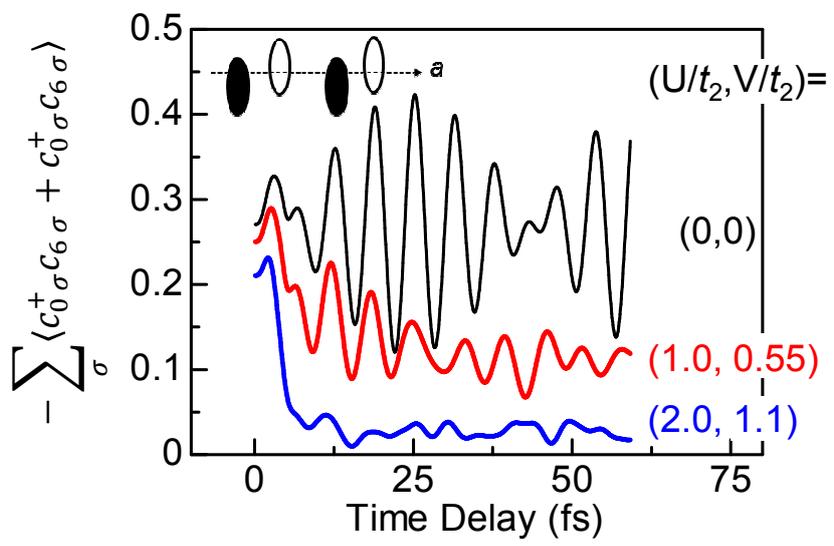

Fig.6

# Supplementary Material for

# "Ultrafast response of plasma-like reflectivity edge in (TMTTF)$_2$AsF$_6$ driven by 7-fs, 1.5-cycle strong-light field"

Y. Naitoh[1], Y. Kawakami[1], T. Ishikawa[1], Y. Sagae[1], H. Itoh[1], K. Yamamoto[2],

T. Sasaki[3], M. Dressel[4], S. Ishihara[1], Y. Tanaka[5], K. Yonemitsu[5], and S. Iwai[1*]

   Here, we provide additional details regarding the temperature dependence of the steady state reflectivity and the spectral analysis using the Drude-Lorentz model. We also show the transient reflectivity $\Delta R$ spectrum instead of $\Delta R/R$ to consider the accuracy of the data when the value of $R$ is very small. Furthermore, the accuracy of the fitting parameter $\Delta\omega_p/\omega_p$ for the transient reflectivity is described to show that the uncertainty is much lower than the field-induced change in $\Delta\omega_p/\omega_p$ of ~3%. The reproducibility of $\Delta R/R$ is also shown.

### Steady state reflectivity and analysis

   Reflectivity spectra for TMTTF salts have been measured and analyzed using the Drude–Lorentz model for >30 years [1-3]. In these compounds, the reflectivity spectrum has an edge that is analogous to the plasma-edge in Drude metals. This reflectivity edge in the near-infrared region has been characterized by $\hbar\omega_p = \sqrt{ne^2/(\varepsilon_\infty\varepsilon_0 m)}$ in the Lorentz model, if $\omega_0$ is much smaller than $\omega_p$ (the number of charges in the 1/4 filling-band $n$ : ~2 x

$10^{21}$ cm$^{-3}$ and their mass $m$: 3~4$m_0$, dielectric constants for high-frequency and vacuum $\varepsilon_\infty, \varepsilon_0$, charge gap $\hbar\omega_0$: ~0.2 eV).

On the other hand, a Raman active C=C vibration becomes infrared active in the low energy (< 0.2 eV) reflectivity spectrum, because of symmetry breaking induced by the electron-molecular vibrational (EMV) coupling. We consider the Fano interference between the electronic transition and this vibrational transition by the dimer model [1]. However, in Fig. 1(b) in the main text, we noticed a disagreement between the observed spectrum and the calculated spectrum for the vibrational region (0.1-0.2 eV) at 25 K, although the disagreement becomes smaller at 150 K. Since such vibrational response at room temperature has been well reproduced previously by this dimer model [1], a possible explanation for the disagreement at low temperatures is the screening of the vibrational peaks by the charge motion driven by the ~0.1 eV AC field of mid infrared light. Here, we describe in further detail the temperature dependence of the steady state reflectivity and the analysis using the Lorentz model, and show that these spectral features at < 0.2 eV do not considerably affect the higher energy region around $\hbar\omega_p$ ~0.7 eV.

Figure s1 shows the reflectivity at 25, 75, 150, and 250 K. On the basis of the dimer model [1], as described above, the reflectivity spectra can be calculated within the framework of Lorentz analysis considering the coupling with intramolecular vibrations as Eqs.(1) – (3).

$$\varepsilon(\omega) = \varepsilon_\infty \left[ 1 + \frac{\omega_p^2}{\omega_0^2 [1 - D(\omega)] - \omega^2 - i\omega\gamma} \right] \quad (1)$$

$$D(\omega) = \sum_{v=1}^{4} \frac{\lambda_v \omega_v^2}{\omega_v^2 - \omega^2 - i\omega\gamma_v} \quad (2)$$

$$\lambda_v = \frac{4\varepsilon_0}{ne^2 d^2} \frac{\omega_p^2}{\omega_0^2} \frac{g_v^2}{\omega_v}, \quad (3)$$

where $\omega_v$ with v=1–4 are the vibrational peak energies, $\lambda_v$ is the dimensionless electron-vibration coupling constant, and $g_v$ is the coupling constant [1], as shown by the solid lines in Fig. s1. Fitting parameters are listed in Table 1. Residual components between the observed and calculated spectra are also indicated by the magenta lines in Fig. s1. For the spectral region > 0.3 eV, the spectra are well fitted at temperatures below 200 K, within the small residual component of ~2%. The disagreement at low temperatures is much smaller than the photoinduced transient reflectivity $\Delta R / R$ >30 % that indicates a decrease in $\omega_p$. On the other hand, the disagreement around $\hbar\omega_p$ ~ 0.7 eV at 250 K is reasonable, considering that the charge coherence becomes worse with increasing temperature. Thus, at temperatures below 200 K, the reflectivity spectrum around $\omega_p$ can be reproduced using the Lorentz model. It is noteworthy that $\gamma$ is almost independent of temperature below $T_{CO}$, and then begins to increase at $T_{CO}$ with increasing temperature. Such an anomaly of $\gamma$ at $T_{CO}$ indicates that the $\gamma$ is governed by an electron–electron scattering process.

Figure s2 shows the reflectivity spectra in the low energy (< 0.25 eV) region at 25 K (a) and at 250 K (b) with the fitting curves. Interestingly, the vibrational peaks are clearer at 250 K than at 25 K. Such suppression of the vibrational peaks at low temperatures is in contrast to the well-known temperature dependence governed by phonon–phonon and/or electron–phonon scatterings. Furthermore, the vibrational peaks can be better fitted at 250 K than at 25 K, i.e., the disagreement clearly becomes larger at low temperatures for this region. This feature along with the high reflectivity of ~0.75 at < 0.15 eV, suggests that the vibrational responses are screened by the charge motion driven by the AC field with 0.1 -0.2 eV. However, as described in the main text, the parameters $\omega_0, \omega_v, \gamma_v$, and $g$ reflecting the low energy (< 0.2 eV) spectrum, do not considerably affect the spectrum around $\omega_p$ (~0.7 eV), which is justified by the fact that $\omega_0 \ll \omega_p$, $\omega_v \ll \omega_p$. Therefore, we can discuss the spectral changes at higher energy (> 0.5 eV) by the parameters $\omega_p$ and $\gamma$ in Eq. 1.

**Indication of the transient reflectivity by $\Delta R$, instead of $\Delta R/R$.**
In Figs. 2–4 of the main text, we have shown the transient reflectivity spectrum as $\Delta R/R$, which is generally used. However, in this case, the steady state reflectivity at > 0.85 eV is very small, as shown in Fig. 1(b). In that spectral region, $\Delta R/R$ becomes large, even if $\Delta R$ is very small. Here, we discuss the transient reflectivity on the basis of $\Delta R$, instead of $\Delta R/R$. Figures s3(a) and s3(b) show the $\Delta R$ measured at $t_d$ =100 fs after the excitation by the 100-fs pulse and at $t_d$ =18 fs after the excitation by the 7-fs pulse. In the $\Delta R$ spectrum shown in Fig. s3(a), the spectral structure between 0.1 and 0.2 eV becomes more prominent. This can be reproduced by the analysis shown by the red curve. According to this analysis, the reflectivity increase in the low energy region can be explained by the changes in $\omega_p$, $\gamma_p$ and the vibrational parameter $\gamma_{v1}$. On the other hand, the disagreement at higher energies (> 0.9 eV) becomes negligible, as shown in Figs. s3(a) and 3(b).

## Accuracy of the fitting parameter $\Delta\omega_p / \omega_p$

Figures s4(a) and s4(b) show the $\Delta R/R$ measured by 100 fs [s4(a)] and 7 fs [s4(b)] pulses, respectively. The orange line [$\Delta\omega_p / \omega_p = -0.018$, $\Delta\gamma/\gamma = -0.12$ in Fig. s4(a), $\Delta\omega_p / \omega_p = -0.028$ in Fig. s4(b)] are the best-fit curves shown in the main text. If we detune the parameter $\Delta\omega_p / \omega_p$ of about $\pm 0.2\%$, the fitting curves cannot reproduce the results as shown by the green dashed and magenta dashed-dotted lines. Therefore, we can conclude that the fitting error of $\Delta\omega_p / \omega_p$ (about 0.2 %) is sufficiently small to discuss the field-induced $\Delta\omega_p / \omega_p$ of about 3 %.

## Reproducibility of $\Delta R/R$ spectrum

To confirm the reproducibility, we have measured $\Delta R/R$ spectrum in another (TMTTF)$_2$AsF$_6$ sample. As shown in Fig. s5, the characteristics of the $\Delta R/R$ at 20 K, 1.1 mJ/cm² are quite similar to those in Figs. 3 and 4 (at almost the same experimental condition; 15 K, 0.8 mJ/cm²), i.e.,

i) $\Delta R/R$ has a broad peak at 0.8-0.9 eV (shown by the red arrow) during the initial time domain (< 20 fs). Then, the peak moves to the lower-energy side (~0.6 eV) until 80-100 fs (Fig. s5(a)). ii) The spectral dip has been observed at ~0.7 eV ($\sim \omega_p$) in the intermediate time region 30-60 fs (shown by the arrows in Fig. s5(a)). iii) The time-domain oscillation with a period of 20 fs has been observed at >0.8 eV (Fig. s5(b)). By these results, the reproducibility of the transient reflectivity is confirmed at least for what we have mainly discussed about.

Figure captions

Figures s1 Reflectivity spectra of (TMTTF)$_2$AsF$_6$ measured at 25, 75, 150, and 250 K for the polarization $E//a$, where $a$ is the stacking axis of the planar molecules. The solid black lines show the fitting curves calculated by Eqs. 1–3. The residual between the observed and calculated spectra are also shown by the magenta lines. The fitting parameters are listed in Table. 1.

Figure s2 Reflectivity spectra at 25 K (a) and at 250 K (b) in the low energy (< 0.25 eV) region with the fitting curves (solid black lines).

Figure s3 $\Delta R$ measured at $t_d$ =100 fs after the excitation by 100-fs pulse and that at $t_d$ =18 fs after the excitation by 7-fs pulse. The solid lines show the fitting curves calculated by Eqs.1–3 [with the same parameters as those used in Fig. 2(b) and Fig. 4(b) ].

Figure s4 $\Delta R/R$ measured by 100 fs (a) and 7 fs (b) pulses, respectively. The best-fitted curves shown in the main text are indicated by the orange lines. The fitting curves for detuned parameters $\Delta\omega_p/\omega_p$ of ± 0.2% are also shown by the green-dashed and magenta-dashed-dotted lines.

Figure s5 $\Delta R/R$ (a) and time evolution (b) measured by 7 fs pulses in another (TMTTF)$_2$AsF$_6$ sample. These results have been shown to indicate the reproducibility of the $\Delta R/R$. The red and blue arrows indicate the broad peak at $t_d$ =10 fs (red) and the spectral dip at $t_d$ =30, 60 fs (blue). The experimental

conditions (20 K, 1.1 mJ/cm$^2$) are quite similar to those in Figs. 3 and 4 (0.8 mJ/cm$^2$).

|  | 25 K | | | 75 K | | | 150 K | | | 250 K | | |
|---|---|---|---|---|---|---|---|---|---|---|---|---|
| $\hbar\omega_p$ (eV) | 0.703 | | | 0.703 | | | 0.703 | | | 0.703 | | |
| $\hbar\omega_{CT}$ (eV) | 0.180 | | | 0.184 | | | 0.193 | | | 0.242 | | |
| $\hbar\gamma_e$ (eV) | 0.125 | | | 0.126 | | | 0.154 | | | 0.208 | | |
| $\varepsilon_\infty$ | 2.46 | | | 2.46 | | | 2.46 | | | 2.46 | | |
| $v$ | $\hbar\omega_v$ (eV) | $g_v$ (eV) | $\hbar\gamma_v$ (eV) | $\hbar\omega_v$ (eV) | $g_v$ (eV) | $\hbar\gamma_v$ (eV) | $\hbar\omega_v$ (eV) | $g_v$ (eV) | $\hbar\gamma_v$ (eV) | $\hbar\omega_v$ (eV) | $g_v$ (eV) | $\hbar\gamma_v$ (eV) |
| 1 | 0.165 | 0.0294 | 0.0081 | 0.165 | 0.0294 | 0.0081 | 0.166 | 0.0294 | 0.0081 | 0.168 | 0.0294 | 0.0081 |
| 2 | 0.135 | 0.0035 | 0.00088 | 0.135 | 0.0035 | 0.00088 | 0.135 | 0.0035 | 0.00088 | 0.135 | 0.0035 | 0.00088 |
| 3 | 0.114 | 0.0092 | 0.0023 | 0.114 | 0.0092 | 0.0023 | 0.115 | 0.0092 | 0.0023 | 0.115 | 0.0092 | 0.0023 |
| 4 | 0.056 | 0.0220 | 0.0056 | 0.056 | 0.0220 | 0.0056 | 0.056 | 0.0220 | 0.0056 | 0.056 | 0.0220 | 0.0058 |

Naitoh et al. Table s1

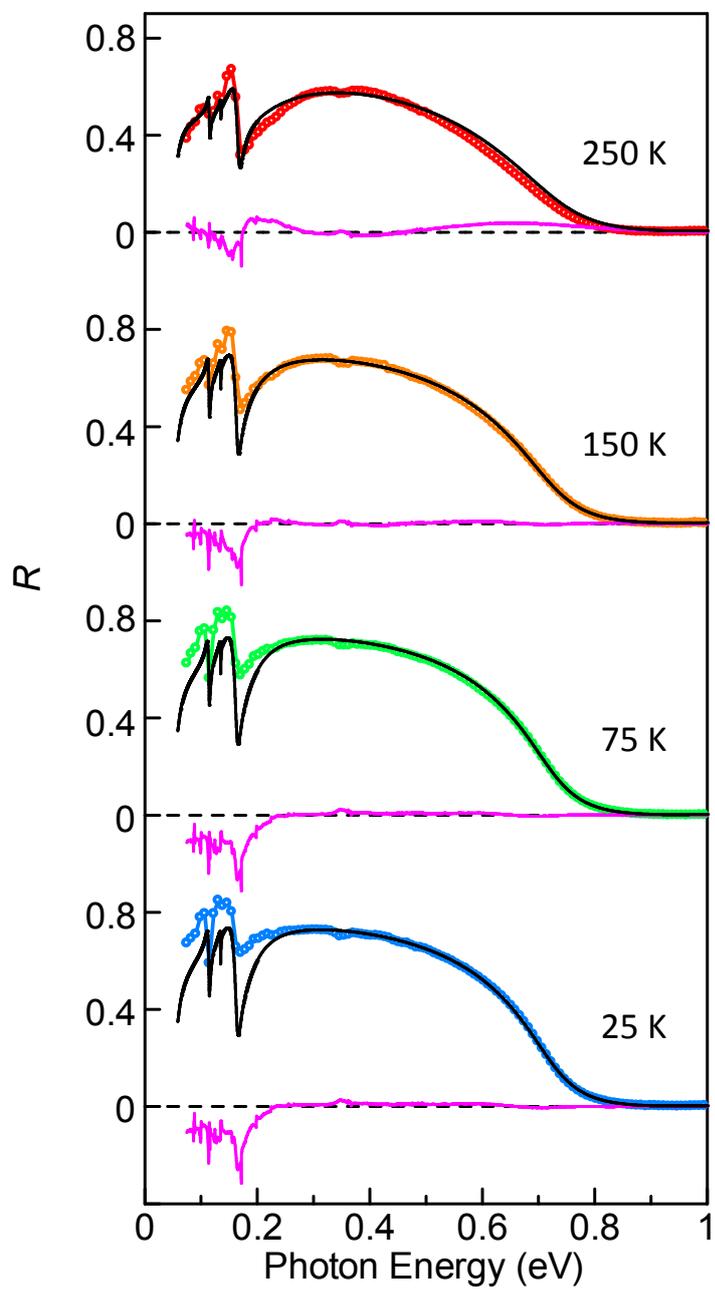

Supplementary Fig. s1

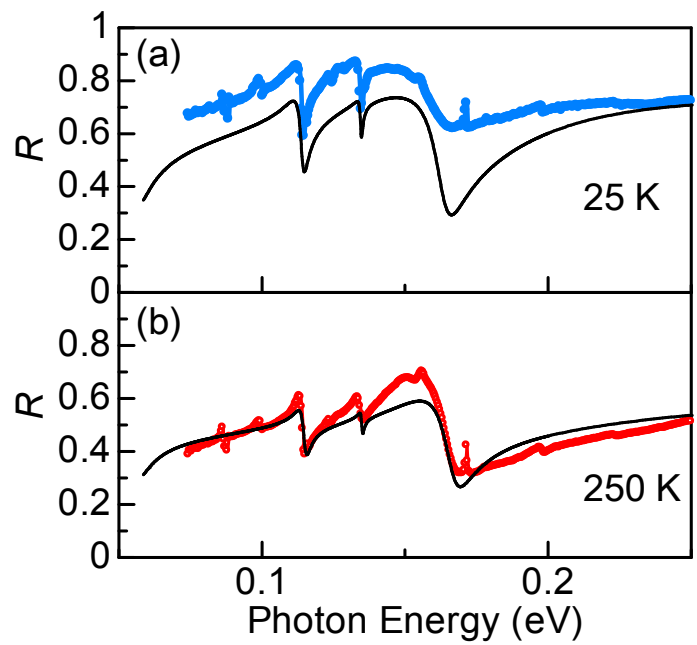

Supplementary Fig. s2

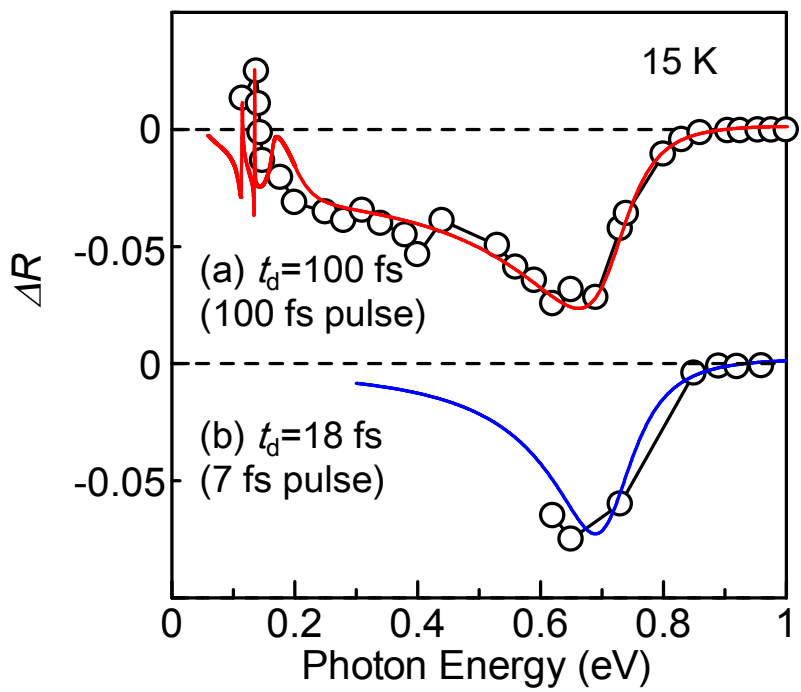

Supplementary Fig.s3

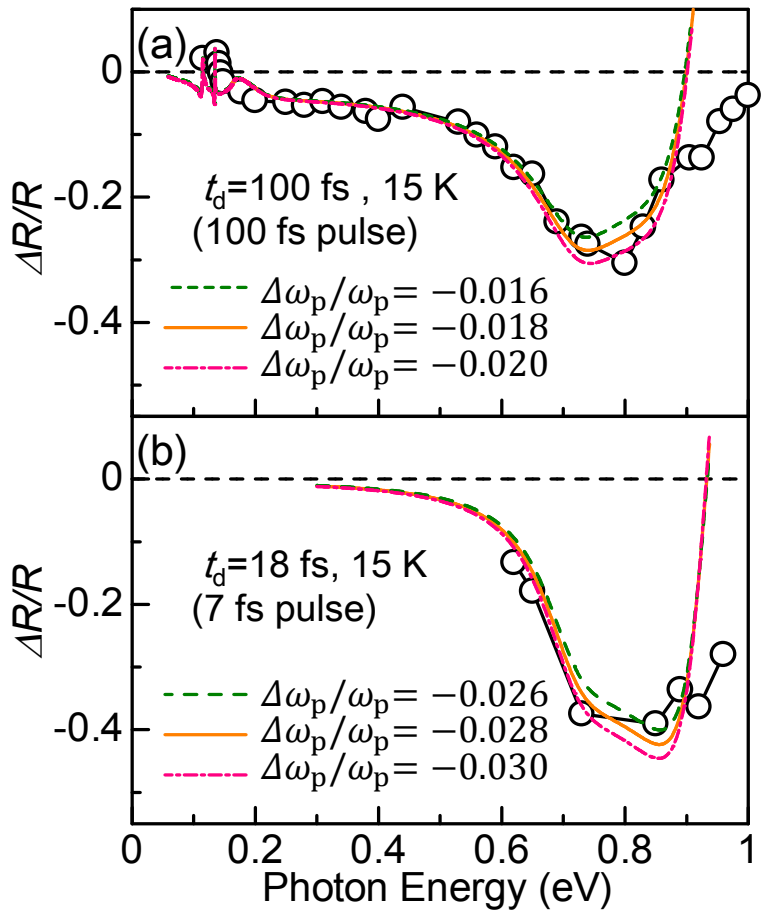

Supplementary Fig.s4

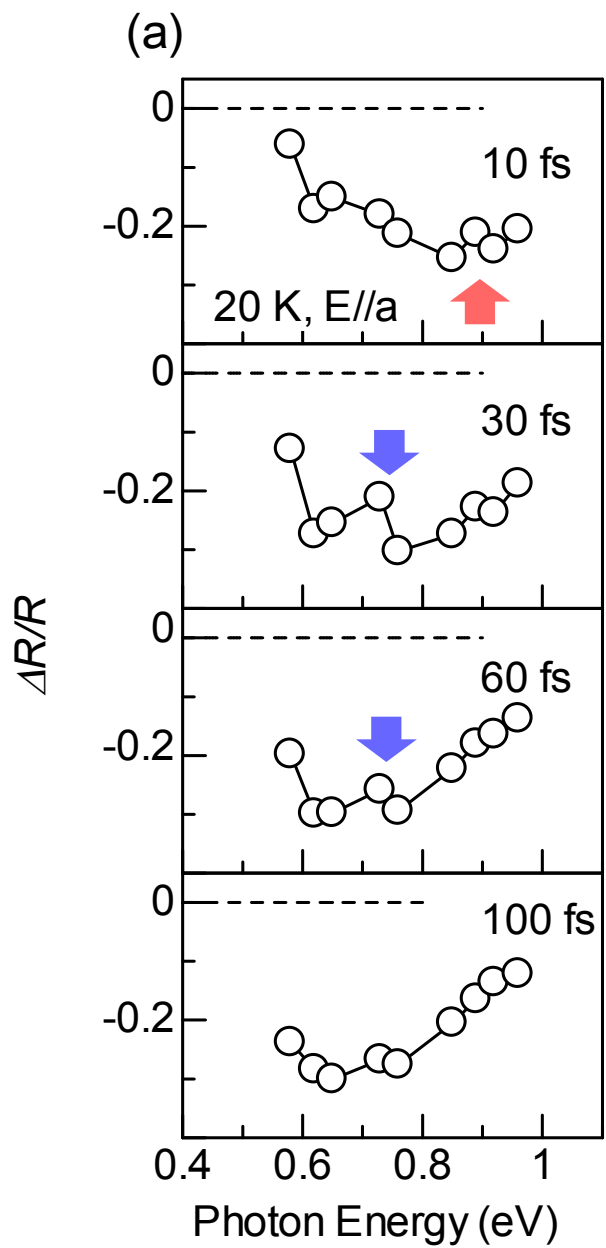
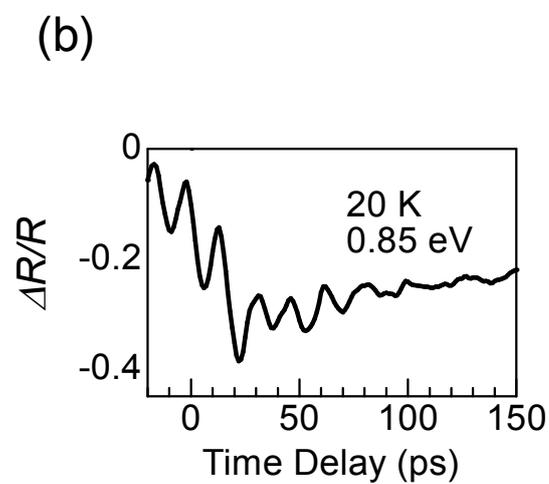

Supplementary Fig.s5